\documentclass{article}
\usepackage{amsmath,amssymb,amsthm,graphicx,setspace,xcolor, bm, bbm}
\usepackage{booktabs}
\usepackage{enumitem}
\usepackage{algpseudocode}
\usepackage{algorithm}
\usepackage{caption}
\usepackage[margin=1in]{geometry}
\usepackage[longnamesfirst]{natbib}
\begin{document}

\title{Fast Test Inversion for Resampling Methods}
\author{Ian Xu\thanks{Department of Economics, Cornell University. E-mail: ianxu.econ@gmail.com}\\ Department of Economics\\ Cornell University}
\maketitle

\begin{abstract}
Randomization-based inference commonly relies on grid search methods to construct confidence intervals by inverting hypothesis tests over a range of parameter values. While straightforward, this approach is computationally intensive and can yield conservative intervals due to discretization. We propose a novel method that exploits the algebraic structure of a broad class of test statistics—including those with variance estimators dependent on the null hypothesis—to produce exact confidence intervals efficiently. By expressing randomization statistics as rational functions of the parameter of interest, we analytically identify critical values where the test statistic's rank changes relative to the randomization distribution. This characterization allows us to derive the exact $p$-value curve and construct precise confidence intervals without exhaustive computation.

For cases where the parameter of interest is a vector and a confidence region is needed, our method extends by calculating and storing the coefficients of the polynomial functions involved. This approach enables us to compute approximate $p$-value functions and confidence regions more efficiently than traditional grid search methods, as we avoid recalculating test statistics from scratch for each parameter value. We illustrate our method using tests from \cite{Pouliot2024} and extend it to other randomization tests, such as those developed by \cite{diciccio:2017aa} and \cite{d2024robust}. Our approach significantly reduces computational burden and overcomes the limitations of traditional grid search methods, providing a practical and efficient solution for confidence interval and region construction in randomization-based inference.\end{abstract}

\newpage
\section{Introduction} \label{intro}

The usual method to produce confidence intervals in randomization-based inference is grid search \citep{garthewaite1996}. This method exploits the duality between hypothesis testing and confidence intervals. Given a level $\alpha$ test, the confidence interval is the set of $\beta_0$ values for which we fail to reject the null hypothesis. While there are times when grid search is appropriate, it should not be our primary method as it has two major drawbacks. 

First, it is computationally inefficient \citep{hansenprob}. For example, in \cite{d2024robust}, the authors develop a randomization test and apply their method to construct confidence intervals for two different examples. In the first example, they construct numerous confidence intervals using different methods by inverting ``the tests of $\beta = \beta_0$ for $\beta_0 \in \{-1.7, -1.69, \dots, 0.3\}$,” resulting in 201 tests. Their method takes 5 seconds for this simple example, which is a considerable amount of time given that there are only 51 observations. In a second example, \cite{d2024robust} “invert the tests of $\beta = \beta_0$ for $\beta_0 \in \{-40, -39.99, \dots, 39.99, 40\}$,” resulting in a staggering 8,001 hypothesis tests. This computational burden becomes even more pronounced for the randomization test developed by \cite{diciccio:2017aa}, which requires approximately 17 hours to complete.\footnote{The supplementary materials estimates that approximately 33 hours are needed altogether to construct the confidence intervals in Table 3, so 17 hours may be conservative.}; this method takes much longer in part because they need to construct a confidence region in $\mathbb{R}^2$ and project onto one dimension, meaning that they need two dimensional grid search. Clearly, there is a need for a faster method to create confidence intervals and regions.

Second, the confidence intervals produced by grid search are conservative. This conservatism arises because the sequence of $\beta_0$ values increases in steps of $\epsilon$, leading to undercoverage if we include only the convex hull of $\beta_0$ values for which we fail to reject. Conversely, if we include adjacent $\beta_0$ values where rejection occurred, the confidence interval will be conservative. One possible alternative is to use bisection search algorithms paired with linear interpolation, but these usually result in approximations \citep{garthewaite1996}.

We develop a method to construct fast and exact confidence intervals by exploiting the structure of the test and randomization statistics. We define exact confidence intervals as confidence intervals that only include the values of $\beta_0$ for which we fail to reject the null hypothesis. To the best of our knowledge, \cite{CaiCanayKimShaikh:2023aa} and \cite{guan2024} present the first exact confidence intervals that use this approach. We extend their work by considering a much broader class of test statistics, including those where the variance estimator depends on the null hypothesis.

As a proof of concept, we primarily use the tests from \cite{Pouliot2024}. He develops a test that can handle general design matrices and provides exact inference under homoskedasticity in small samples and asymptotically exact inference under general model misspecification. We will extend the principles applied to \cite{Pouliot2024} to other randomization tests—including that developed by \cite{d2024robust} and \cite{diciccio:2017aa}—in Section \ref{otherTests} and the Appendix. Since we generally adopt the notation from each source, \textit{readers should consider the notation from each test and its analysis independently unless we explicitly indicate otherwise}.

\section{Working Example: The Exact $t$-Test}

\subsection{Randomization Statistic Formulation}\label{onesidedPouliot}

We encourage readers to read \cite{Pouliot2024}, but to provide some background and perspective, consider the linear regression model
$$
\mathbf{Y} = \mathbf{1}_n \beta_0 + \mathbf{X}_1 \beta_1 + \mathbf{X}_2 \beta_2 + \varepsilon,
$$ 
where $\mathbf{Y}, \mathbf{1}_n, \mathbf{X}_1, \varepsilon \in \mathbb{R}^{n \times 1}$;  $\mathbf{X}_2 \in \mathbb{R}^{n \times p}$; $\beta_0, \beta_1 \in \mathbb{R}$; and $\beta_2 \in \mathbb{R}^{p \times 1}$. 

To test $H_0:\beta_1 = \beta_1^0$ against $H_1:\beta_1 < \beta_1^0$ (see Appendix \ref{pouliotTwoSided} for the two-sided hypothesis test), we use a randomization test by comparing the test statistic against the randomization statistics. The randomization statistic is defined as
$$
T_g(\beta_1^0) \equiv \frac{(\mathcal{Q}_1 \mathbf{X}_1)^\top g (\mathbf{Y} - \mathbf{X}_1 \beta_1^0)}{\hat{\sigma}_g},
$$
where 
$$
\hat{\sigma}_g^2 \equiv \frac{1}{n}\sum_{i=1}^{n} (\mathcal{Q}_1 \mathbf{X}_1)_i^2 (\mathcal{Q}_{2} g \mathbf{Y})_i^2;
$$
$g$ is a block permutation matrix in $\mathbf{G}$, which is the set of all block permutation matrices for a prespecified number of blocks; $\mathcal{Q}_1$ is the annihilator matrix that projects onto the orthogonal complement of the span of $\left\{g\mathbf{X}_2\right\}_{g \in \mathbf{G}}$; and $\mathcal{Q}_2$ is the annihilator matrix that projects onto the orthogonal complement of the span of $\mathbf{1}$, $\left\{g\mathbf{X}_1\right\}_{g \in \mathbf{G}}$, and $\left\{g\mathbf{X}_2\right\}_{g \in \mathbf{G}}$. Let $M = |\mathbf{G}|$, the total number of block permutations $g$ (this definition applies throughout the paper). The test statistic corresponds to $g = \text{Id}$, the identity matrix. 

The full procedure to perform the hypothesis test is detailed in Algorithm \ref{PouliotTestAlg}. We can also obtain a $p$-value for a given $\beta_1^0$ value by calculating the proportion of randomization statistics that are greater or equal to the test statistic. This is essentially comparing the rank of the test statistic with the randomization distribution.

\begin{algorithm}
\caption{Algorithm to test right-sided alternative using \cite{Pouliot2024}}
\label{PouliotTestAlg}
\begin{algorithmic}[1]
    \State \textbf{Initialize:} Empty list for randomization statistics \hfill \textit{O(1)}
    \State Calculate test statistic $T_{\text{Id}}(\beta_1^0)$ \hfill \textit{O(n)}
    \State Add $T_{\text{Id}}(\beta_1^0)$ to list of randomization statistics \hfill \textit{O(1)}
    \For{$g$ in $\mathbf{G} \setminus \text{Id}$} \hfill \textit{O(M)}
        \State Calculate randomization statistic $T_{g}(\beta_1^0)$ \hfill \textit{O(n)}
        \State Add $T_{g}(\beta_1^0)$ to list of randomization statistics \hfill \textit{O(1)}
    \EndFor
    \State Calculate $R \gets 1-\alpha$ quantile of the randomization statistics \hfill \textit{O(M log M)}
    \If{$T_{\text{Id}}(\beta_1^0) > R$} \hfill \textit{O(1)}
        \State \textbf{Output:} Reject $H_0: \beta_1 = \beta_1^0$ \hfill \textit{O(1)}
    \Else
        \State \textbf{Output:} Fail to Reject $H_0: \beta_1 = \beta_1^0$ \hfill \textit{O(1)}
    \EndIf
\end{algorithmic}
To test the left-sided alternative, follow the same procedure except $R$ will be the $\alpha$ quantile—using supremum in lieu of infimum in the definition of quantile—of the randomization statistics and we reject $H_0$ if and only if $T_{\text{Id}}(\beta_1^0) < R$.
\end{algorithm}

\subsection{Constructing Confidence Intervals}

As noted above, the traditional method to construct confidence intervals is grid search, but this method is computationally inefficient and results in conservative confidence intervals. There are two other challenges with this method here. First, we don't know where the center of the confidence interval is, so we don't know the range of $\beta_1^0$ values to search over. Traditionally, we would create a grid of $\beta_1^0$ values to the left and right of $\hat{\beta}_1$ because the confidence intervals are centered at $\hat{\beta}_1$, but $\hat{\beta}_1$ may not be the center of the interval here.\footnote{Even creating a grid of $\beta_1^0$ values from $\hat{\beta}_1 - 4\text{s.e.}(\hat{\beta}_1)$ to $\hat{\beta}_1 + 4\text{s.e.}(\hat{\beta}_1)$ will not necessarily solve this issue because the confidence intervals can be wide for poor orderings of the data that reduce power (i.e. $\mathcal{Q}_1$ orthogonalizes almost all variation in $\mathbf{X}_1$.} Second, we don't know when to stop searching. The confidence interval from randomization tests can have ``holes''—the interval could be a union of two disjoint sets. This arises because the $p$-value function may not be monotonic, so it could be the case that the $p$-value initially decreases as $\beta_1^0$ gets further from the truth, but later increases. 

\begin{figure}[h]
    \centering
    \includegraphics[width=1\linewidth]{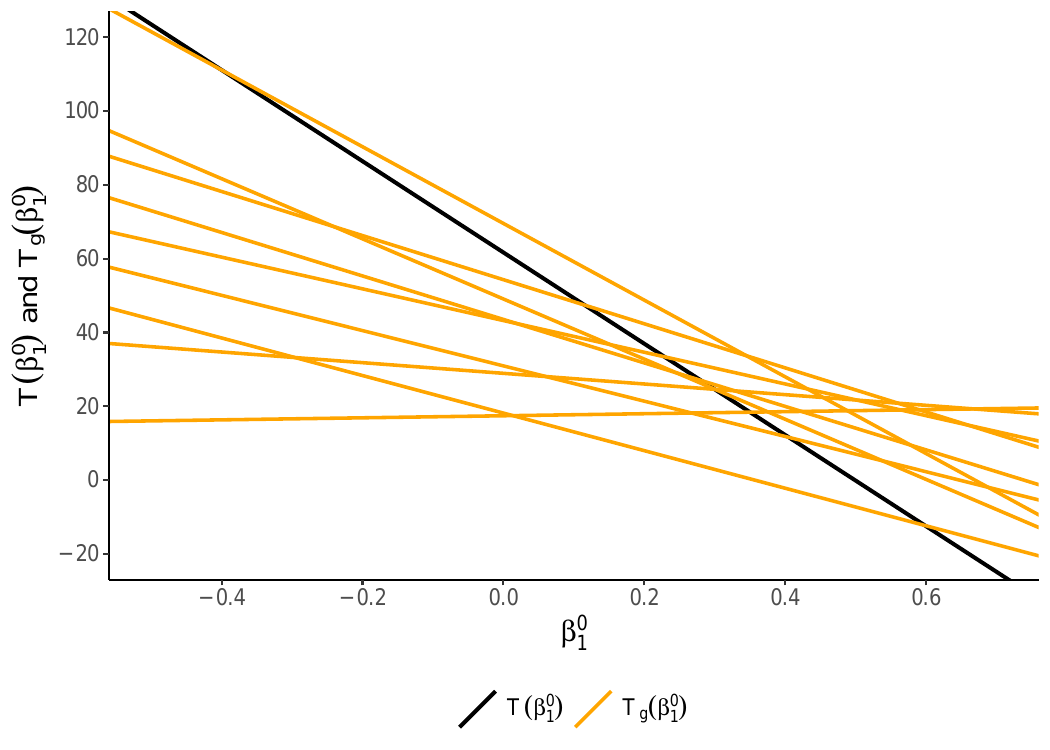}
    \caption{We plot $T(\beta_1^0)$ and $T_g(\beta_1^0)$ for 9 permutations using data from Example 3.1 of \cite{Wooldridge:2020aa}.}
    \label{fig:testrandlines}
\end{figure}

The solution to these woes is to pair the structure of the test statistic with algebra and geometry to find the exact $p$-value function from which we can derive precise confidence intervals for any choice of $\alpha$. To see this, note that the randomization statistics can be reformulated as:
$$
\begin{aligned}
T_g(\beta_1^0) &= \frac{(\mathcal{Q}_1 \mathbf{X}_1)^\top g \mathbf{Y}}{\hat{\sigma}_g} - \frac{(\mathcal{Q}_1 \mathbf{X}_1)^\top g \mathbf{X}_1}{\hat{\sigma}_g} \beta_1^0 \\
&= m_g\beta_1^0 + b_g,
\end{aligned}
$$
where $m_g = -(\mathcal{Q}_1 \mathbf{X}_1)^\top g \mathbf{X}_1 / \hat{\sigma}_g$ is the slope and $b_g = (\mathcal{Q}_1 \mathbf{X}_1)^\top g \mathbf{Y} / \hat{\sigma}_g$ is the intercept for the randomization statistic with block permutation $g$. From this, we see that randomization statistics (including the test statistic) are lines. This means that each of the randomization statistics $T_g(\beta_1^0)$ (excluding the test statistic) will be exactly equal to $T_\text{Id}(\beta_1^0)$ at a unique $\beta_1^0$ value because two lines with differing slopes must intersect at a single point.\footnote{Two lines can intersect at one, zero or an infinite number of points, but the latter two require the slopes to be equal, which has zero probability.} We denote these $\beta_1^0$ values for which the test statistic line intersects the randomization statistic line as $\dot{\beta}_1^{0, g}$. Finding these $\dot{\beta}_1^{0, g}$ values only requires a bit of algebra; we just need to set the test statistic equal to the randomization statistic and solve for $\beta_1^0$:
$$
\begin{aligned}
    T_{\text{Id}}(\beta_1^0) &= T_g(\beta_1^0) \\
    \Leftrightarrow m_{\text{Id}}\beta_1^0 + b_{\text{Id}} &= m_g \beta_1^0 + b_g \\
    \Leftrightarrow m_{\text{Id}}\beta_1^0 - m_g \beta_1^0 &= b_g - b_{\text{Id}} \\
    \Rightarrow \dot{\beta}_1^{0, g} &= \frac{b_g - b_{\text{Id}}}{ m_{\text{Id}} - m_g},
\end{aligned}
$$
where $m_{\text{Id}} \neq m_g$. With $\dot{\beta}_1^{0, g}$ and the slope of the test and randomization statistics, we can easily deduce the relationship between $T_{\text{Id}}(\beta_1^0)$ and any randomization statistic $T_g(\beta_1^0)$ for all $\beta_1^0 \in \mathbb{R}$.\footnote{If $m_{\text{Id}} = m_g$, then  $b_{\text{Id}} > b_g \Leftrightarrow T_{\text{Id}} > T_g$.} For example, notice that if $m_{\text{Id}} < m_g$, then $T_{\text{Id}} > T_g$ for all $\beta_1^0 < \dot{\beta}_1^{0, g}$, and $T_{\text{Id}} < T_g$ for all $\beta_1^0 > \dot{\beta}_1^{0, g}$. These $\dot{\beta}_1^{0, g}$ values are critical because it is only at these specific $\beta_1^0$ values that the rank of the test statistic changes in relation to the randomization distribution. At all other $\beta_1^0$ values, the test and randomization statistics are changing in value, but the rank of the test statistic does not change in relation to the randomization distribution, meaning that the $p$-value does not change. See Figure \ref{fig:testrandlines} for a visualization.

\begin{figure}[h]
    \centering
    \includegraphics[width=1\linewidth]{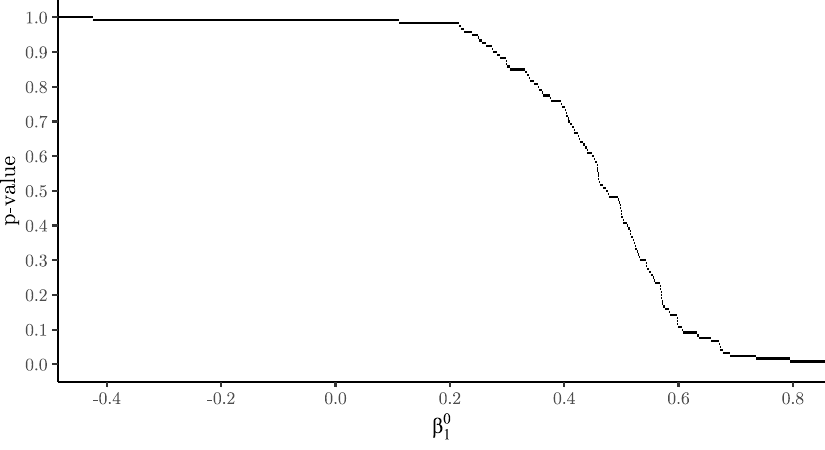}
    \caption{This is the exact $p$-value curve for hsGPA in Example 3.1 of \cite{Wooldridge:2020aa} using the test developed by \cite{Pouliot2024} with 120 total permutations, including the identity permutation.}
    \label{fig:pvalcurve}
\end{figure}

The $p$-value function is 
$$
p(\beta_1^0) \equiv \frac{1}{M}\left(1 + \sum_{g \in \mathbf{G}_n \setminus \text{Id}} \mathbbm{1}[T_{\text{Id}}(\beta_1^0) \leq T_g(\beta_1^0)]\right).
$$
Note that the $p$-value curve is a step function (see Figure \ref{fig:pvalcurve} for an example). We can find the curve using Algorithm \ref{pvalCurve1}. The exact level $1-\alpha$ confidence interval is the set of $\beta_1^0$ values with $p$-value greater than $\alpha$, which we can find by filtering the rows of $F$, which is the output of Algorithm \ref{pvalCurve1}.

\begin{algorithm}
\caption{Algorithm to compute $p$-value curve for right-sided hypothesis test using \cite{Pouliot2024}}
\label{pvalCurve1}
\begin{algorithmic}[1]
    \State \textbf{Initialize:} Empty list $W$ to store line parameters and intersections: $\left(m_g, \ b_g, \ \dot{\beta}_1^{0,g}\right)$ \hfill \textit{O(1)}
    \State \textbf{Initialize:} Empty list $F$ to store intervals and $p$-values: $(\beta_1^{\text{start}}, \ \beta_1^{\text{end}}, \ p\text{-value})$ \hfill \textit{O(1)}
    \State \textbf{Initialize:} $\text{Counter} \gets 1$
    \hfill \textit{O(1)}
    \State Calculate $m_{\text{Id}}$ and $b_{\text{Id}}$ \hfill \textit{O(n)}
    \For{$g$ in $\mathbf{G} \setminus \text{Id}$} \hfill \textit{O(M)}
        \State Calculate $m_g$ and $b_g$ \hfill \textit{O(n)}
        \If{$m_{\text{Id}} \neq m_g$}
            \State Compute $\dot{\beta}_1^{0,g}$ \hfill \textit{O(1)}
            \State Append $\left(m_g, b_g, \dot{\beta}_1^{0,g}\right)$ to $W$ \hfill \textit{O(1)}
        \ElsIf{$b_{\text{Id}} > b_g$}
            \State $\text{Counter} \gets \text{Counter} + 1$ \hfill \textit{O(1)}
        \EndIf
    \EndFor
    \State Sort $W$ by increasing $\dot{\beta}_1^{0,g}$ \hfill \textit{O(M log M)}
    \State Append $\left(-\infty,\ W[1].\dot{\beta}_1^{0,g},\ \text{Counter}\right)$ to $F$ \hfill \textit{O(1)}
    \For{$i$ from $1$ to $M-1$}
        \If{$m_{\text{Id}} < W[i].m_g$}
            \State $\text{Counter} \gets \text{Counter} + 1$ \hfill \textit{O(1)}
        \Else
            \State $\text{Counter} \gets \text{Counter} - 1$ \hfill \textit{O(1)}
        \EndIf
        \If{$i < M$}
            \State Append $\left( W[i].\dot{\beta}_1^{0,g},\ W[i+1].\dot{\beta}_1^{0,g},\ \text{Counter} \right)$ to $F$ \hfill \textit{O(1)}
        \Else
            \State Append $\left( W[i].\dot{\beta}_1^{0,g},\ \infty,\ \text{Counter} \right)$ to $F$ \hfill \textit{O(1)}
        \EndIf
    \EndFor
    \State \textbf{Output:} $F$, which represents the $p$-value curve (a step function)
\end{algorithmic}
Since duplicate values of $\dot{\beta}_1^{0,g}$ may occur with measure zero, filter out all rows of $F$ where $\beta_1^{\text{start}} = \beta_1^{\text{end}}$.
\end{algorithm}

We also make the observation here that $T_g(\beta_1^0) \in \mathbb{R}[\beta_1^0]$, a scalar polynomial of degree 1 over the field $\mathbb{R}$. This perspective does not necessarily unlock new insights in this simple example, but we nevertheless introduce it here as it will be helpful with more complex randomization statistics like the over-identified instrumental variable case discussed in the following section.

\section{Extension to Wald-like Statistics} \label{waldPouliot}

\cite{Pouliot2024} also defines a test for overidentified instrumental variable analysis. Suppose there are $k$ instruments and $\beta_1^0$ is a scalar. Then, the heteroskedasticity-robust Wald-like randomization statistic is
$$
t_g(\beta_1^0) = T^{iv}_g(\beta_1^0)^T \hat{\Sigma}_g(\beta_1^0)^{-1} T^{iv}_g(\beta_1^0),
$$
where
$$
\begin{aligned}
    T^{iv}_g(\beta_1^0) &= (\mathcal{Q}_1 \mathbf{Z})^T g (\mathbf{Y} - \mathbf{X}_1 \beta_1^0); \\
    \hat{\Sigma}_g(\beta_1^0) &= \frac{1}{n}\sum_{i=1}^{n} (\mathcal{Q}_1 \mathbf{Z})_i (\mathcal{Q}_1 \mathbf{Z})_i^T \left(\mathcal{Q}_3 g (\mathbf{Y} - \mathbf{X}_1 \beta_1^0)\right)_i^2;
\end{aligned}
$$
$\mathbf{Y}, \mathbf{X}_1 \in \mathbb{R}^{n \times 1}$; $\mathbf{X}_2 \in \mathbb{R}^{n \times p}$; $\mathbf{Z}_2 \in \mathbb{R}^{n \times k}$; $\mathcal{Q}_1$ is defined in Section \ref{onesidedPouliot}; and $\mathcal{Q}_3$ is the annihilator matrix that projects onto the orthogonal complement of the span of $\mathbf{1}$, $\mathbf{X}_1$, and $\left\{g\mathbf{X}_2\right\}_{g \in \mathbf{G}}$. Note that unlike the previous example where the variance estimator $\hat{\sigma}_g^2$ was not a function of $\beta_1^0$, that is not the case here with $\hat{\Sigma}_g(\beta_1^0)$.\footnote{This statistic slightly differs from that in the paper as the variance estimator $\hat{\Sigma}_g$ is a function of $\beta_1^0$. This variance estimator slightly increases power.}

We would like to apply the same principles as before: transform the test and randomization statistics into some form where we can find the $\beta_1^0$ values for which the test statistic intersects the randomization statistic. The issue here is that our covariance matrix estimator—that we need to invert—is sandwiched between two outer "bread" matrices. We cannot set the test and randomization statistics equal to each other and multiply $\hat{\Sigma}_g(\beta_1^0)$ and $\hat{\Sigma}_\text{Id}(\beta_1^0)$ on both sides to remove $\hat{\Sigma}_g(\beta_1^0)^{-1}$ and $\hat{\Sigma}_\text{Id}(\beta_1^0)^{-1}$. We solve this problem by recognizing that $\hat{\Sigma}_g(\beta_1^0)$ is in $\mathbb{R}^{k \times k}[\beta_1^0]$: it is a symmetric polynomial matrix where each element is a quadratic function in $\beta_1^0$. To see this, observe that $\hat{\Sigma}_g(\beta_1^0)$ can be reformulated as
$$
\renewcommand{\arraystretch}{1.25} 
\hat{\Sigma}_g(\beta_1^0) = \begin{bmatrix}
    \hat{s}_g^{11}(\beta_1^0) & \hat{s}_g^{12}(\beta_1^0) & \dots & \hat{s}_g^{1k}(\beta_1^0) \\
    \hat{s}_g^{21}(\beta_1^0) & \hat{s}_g^{22}(\beta_1^0) & \dots & \hat{s}_g^{2k}(\beta_1^0) \\
    \vdots & \vdots & \ddots & \vdots \\
    \hat{s}_g^{k1}(\beta_1^0) & \hat{s}_g^{k2}(\beta_1^0) & \dots & \hat{s}_g^{kk}(\beta_1^0)
\end{bmatrix},
$$
where for $1 \leq v,w \leq k$,
$$
\begin{aligned}
\hat{s}_g^{vw}(\beta_1^0) &= \frac{1}{n}\sum_{i=1}^{n} (\mathcal{Q}_1 \mathbf{Z}_v)_i (\mathcal{Q}_1 \mathbf{Z}_w)_i \left(\mathcal{Q}_3 g (\mathbf{Y} - \mathbf{X}_1 \beta_1^0)\right)_i^2 \\
&= \frac{1}{n} \left(\mathcal{Q}_3 g (\mathbf{Y} - \mathbf{X}_1 \beta_1^0)\right)^T \text{diag}[\mathcal{Q}_1 \mathbf{Z}_v \circ \mathcal{Q}_1 \mathbf{Z}_w] \left(\mathcal{Q}_3 g (\mathbf{Y} - \mathbf{X}_1 \beta_1^0)\right) \\
&= a_g^{vw} + b_g^{vw}(\beta_1^0) + c_g^{vw}(\beta_1^0)^2.
\end{aligned}
$$
Here, diag$[\bullet]$ denotes the matrix formed from vector $\bullet$, and $\circ$ denotes the Hadamard (element-wise) product (these definitions apply throughout the paper). Then, 
$$
\begin{aligned}
a_g^{vw} &= \mathbf{Y}^T g^T \mathcal{Q}_3 \text{diag}[(\mathcal{Q}_1 \mathbf{Z}_{v}) \circ (\mathcal{Q}_1 \mathbf{Z}_{w})] \mathcal{Q}_3 g \mathbf{Y}, \\
b_g^{vw} &= -2\mathbf{X}_1^T g^T \mathcal{Q}_3 \text{diag}[(\mathcal{Q}_1 \mathbf{Z}_{v}) \circ (\mathcal{Q}_1 \mathbf{Z}_{w})] \mathcal{Q}_3 g\mathbf{Y}, \\
c_g^{vw} &= \mathbf{X}_1^T g^T \mathcal{Q}_3 \text{diag}[(\mathcal{Q}_1 \mathbf{Z}_{v}) \circ (\mathcal{Q}_1 \mathbf{Z}_{w})] \mathcal{Q}_3 g\mathbf{X}_1.
\end{aligned}
$$

$\hat{\Sigma}_g(\beta_1^0)$ is invertible if the determinant is a non-zero polynomial \citep{kaczorek2007polynomial}. This means that a sufficient condition for $\hat{\Sigma}_g(\beta_1^0)$ to be invertible is if it is invertible when $\beta_1^0 = 0$ (i.e. the determinant of $\hat{\Sigma}_g(0) \neq 0$), which should occur with probability one. We can invert $\hat{\Sigma}_g(\beta_1^0)$ using cofactor expansion. Thus, our Wald-like statistic can be rewritten as
$$
\begin{aligned}
    t_g(\beta_1^0) &= T^{iv}_g(\beta_1^0)^T \hat{\Sigma}_g(\beta_1^0)^{-1} T^{iv}_g(\beta_1^0) \\
    &= \frac{T^{iv}_g(\beta_1^0)^T \text{adj}[\hat{\Sigma}_g(\beta_1^0)] T^{iv}_g(\beta_1^0)}{\text{det}[\hat{\Sigma}_g(\beta_1^0)]} \\
    &= \frac{N_g(\beta_1^0)}{D_g(\beta_1^0)},
\end{aligned}
$$
where $N_g(\beta_1^0) = T^{iv}_g(\beta_1^0)^T \text{adj}[\hat{\Sigma}_g(\beta_1^0)] T^{iv}_g(\beta_1^0)$, and $D_g(\beta_1^0) = \text{det}[\hat{\Sigma}_g(\beta_1^0)]$. Here, $\text{adj}[\bullet]$ and $\text{det}[\bullet]$ denote the adjugate and determinant of $\bullet$ respectively (these definitions apply throughout the paper). Note that $T^{iv}_g(\beta_1^0) \in \mathbb{R}^{k \times 1}[\beta_1^0]$, and $\text{adj}[\hat{\Sigma}_g(\beta_1^0)] \in \mathbb{R}^{k \times k}[\beta_1^0]$, so $N_g(\beta_1^0)$ and $D_g(\beta_1^0)$ are scalars in $\mathbb{R}[\beta_1^0]$. 

These Wald-like statistics $t_g(\beta_1^0)$ differ from the statistics in the previous section because $t_g(\beta_1^0)$ is a rational function. Regardless, we can still find the intersection values algebraically by finding the real roots of $D_g(\beta_1^0)N_{\text{Id}}(\beta_1^0) - D_{\text{Id}}(\beta_1^0)N_g(\beta_1^0)$ = 0; we just need to ensure that $D_{\text{Id}}(\beta_1^0)$ and $D_g(\beta_1^0)$ are non-zero for these solutions. After finding the intersection values, we can use an algorithm similar to Algorithm \ref{PouliotTestAlg} to find the exact $p$-value curve from which we can easily extract the level $1-\alpha$ confidence interval(s); filter the output to only include rows for which the $p$-value is greater than $\alpha$, then the confidence interval(s) are the regions defined by $\beta_1^{\text{start}}$ and $\beta_1^{\text{end}}$.

\section{Jointly Valid Confidence Regions} \label{jointValidPouliot}

Thus far we have only considered tests where the coefficient of interest $\beta_1^0$ is a scalar, but economists are often interested in constructing confidence regions for multiple coefficients and projecting this confidence region onto each dimension to create projected confidence intervals. We present two methods to create such intervals—fast grid search and optimization. Fast grid search works generally, but does not yield exact projected confidence intervals. We can use optimization to find the exact projected confidence intervals, but this requires assumptions that we discuss later.

Consider the overidentified instrumental variable statistic from \cite{Pouliot2024}, but where $\beta_1^0 \equiv [\beta^0_{1,1}, \ \beta^0_{1,2}]^T$ $\in \mathbb{R}^{2 \times 1}$ and we annihilate all $g \mathbf{X}_1$ variation when estimating $\hat{\varepsilon}$ so that $\hat{\Sigma}_g$ is not a function of $\beta_1^0$.\footnote{This is the precise Wald-like randomization test in \cite{Pouliot2024}.} In this case, the Wald-like statistic is 
$$
\begin{aligned}
    t_g(\beta_{1,1}^0, \beta_{1,2}^0) &\equiv t_g(\beta_1^0) \\
    &= T_g^{iv}(\beta_1^0)^T \hat{\Sigma}_g^{-1} T_g^{iv}(\beta_1^0)
\end{aligned}
$$
where
$$
\begin{aligned}
    T_g^{iv}(\beta_1^0) &= (\mathcal{Q}_1 \mathbf{Z})^T g (\mathbf{Y} - \mathbf{X}_1 \beta_1^0) \\ 
    &= (\mathcal{Q}_1 \mathbf{Z})^T g (\mathbf{Y} - \mathbf{X}_{1,1} \beta_{1,1}^0 - \mathbf{X}_{1,2} \beta_{1,2}^0) \\
\end{aligned}
$$
and
$$
\hat{\Sigma}_g = \frac{1}{n}\sum_{i=1}^{n} (\mathcal{Q}_1 \mathbf{Z})_i (\mathcal{Q}_1 \mathbf{Z})_i^T \left(\mathcal{Q}_2 g \mathbf{Y}\right)_i^2.
$$
Here, $\mathbf{X}_1 \in \mathbb{R}^{n \times 2}$; and $\mathbf{Y}$, $\mathbf{X}_2$, $\mathbf{Z}$, $\mathcal{Q}_1$, and $\mathcal{Q}_2$ are defined in Section~\ref{onesidedPouliot}. Similar to before, $\hat{\Sigma}_g$ can be rewritten as
$$
\begin{aligned}
\hat{\Sigma}_g &= \begin{bmatrix}
    \hat{s}_g^{11} & \hat{s}_g^{12} & \dots & \hat{s}_g^{1k} \\
    \hat{s}_g^{21} & \hat{s}_g^{22} & \dots & \hat{s}_g^{2k} \\
    \vdots & \vdots & \ddots & \vdots \\
    \hat{s}_g^{k1} & \hat{s}_g^{k2} & \dots & \hat{s}_g^{kk}
\end{bmatrix},
\end{aligned}
$$
where 
$$
\begin{aligned}
\hat{s}_g^{vw} &= \frac{1}{n}\sum_{i=1}^{n} (\mathcal{Q}_1 \mathbf{Z}_v)_i (\mathcal{Q}_1 \mathbf{Z}_w)_i \left(\mathcal{Q}_2 g \mathbf{Y}\right)_i^2 \\
&= \frac{1}{n} \left(\mathcal{Q}_2 g \mathbf{Y}\right)^T \text{diag}[\mathcal{Q}_1 \mathbf{Z}_v \circ \mathcal{Q}_1 \mathbf{Z}_w] \left(\mathcal{Q}_2 g \mathbf{Y}\right).
\end{aligned}
$$
Notice here that $\hat{\Sigma}_g$ is a real valued matrix that we can invert using cofactor expansion (assuming it is invertible). Then $t_g(\beta_1^0)$ can be rewritten
$$
\begin{aligned}
t_g(\beta_{1,1}^0, \beta_{1,2}^0) &= T^{iv}_g(\beta_1^0)^T \hat{\Sigma}_g^{-1} T^{iv}_g(\beta_1^0) \\
    &= \mathbf{X}^T_{1,1} A_g \mathbf{X}_{1,1} (\beta_{1,1}^0)^2 + 2\mathbf{X}^T_{1,1} A_g \mathbf{X}_{1,2}(\beta^0_{1,1} \ \beta^0_{1,2}) + \mathbf{X}^T_{1,2} A_g \mathbf{X}_{1,2} (\beta_{1,2}^0)^2 \\
    & \quad \quad - 2\mathbf{Y}^T A_g \mathbf{X}_{1,1}(\beta^0_{1,1}) - 2\mathbf{Y}^T A_g \mathbf{X}_{1,2}(\beta^0_{1,2}) + \mathbf{Y}^T A_g \mathbf{Y} \\
    &=
\begin{bmatrix}
\beta_{1,1}^0, \ \beta_{1,2}^0, \ 1
\end{bmatrix}
\begin{bmatrix}
\mathbf{X}^T_{1,1} A_g \mathbf{X}_{1,1} & \mathbf{X}^T_{1,1} A_g \mathbf{X}_{1,2} & -\mathbf{Y}^T A_g \mathbf{X}_{1,1} \\
\mathbf{X}^T_{1,1} A_g \mathbf{X}_{1,2} & \mathbf{X}^T_{1,2} A_g \mathbf{X}_{1,2} & -\mathbf{Y}^T A_g \mathbf{X}_{1,2} \\
-\mathbf{Y}^T A_g \mathbf{X}_{1,1} & -\mathbf{Y}^T A_g \mathbf{X}_{1,2} & \mathbf{Y}^T A_g \mathbf{Y}
\end{bmatrix}
\begin{bmatrix}
\beta_{1,1}^0 \\
\beta_{1,2}^0 \\
1
\end{bmatrix} \\
    &= 
\begin{bmatrix}
\beta_{1,1}^0, \ \beta_{1,2}^0, \ 1
\end{bmatrix}
\Omega_g
\begin{bmatrix}
\beta_{1,1}^0 \\
\beta_{1,2}^0 \\
1
\end{bmatrix} \\
\end{aligned}
$$
where $A_g = g^T \mathcal{Q}_1 \mathbf{Z} \hat{\Sigma}_g^{-1} \mathbf{Z}^T \mathcal{Q}_1 g$ and $\Omega_g$ is the center matrix in the third equality. Note that $t_g(\beta_1^0)$ can define a conic section: it is a polynomial of degree two in $\beta_{1,1}^0$ and $\beta_{1,2}^0$.
Like above, we want to consider the values of $[\beta^0_{1,1}, \ \beta^0_{1,2}]^T$ for which $t_{\text{Id}}(\beta_1^0) = t_g(\beta_1^0)$ because it is at these $[\beta_{1,1}^0, \beta_{1,2}^0]^T$ values that the rank of $t_{\text{Id}}(\beta_1^0)$ changes relative to the distribution of $t_g(\beta_1^0)$, which changes the $p$-value. To find this set of $[\beta_{1,1}^0, \beta_{1,2}^0]^T$ values, we adopt a similar procedure to that above:
$$
\begin{aligned}
    t_{\text{Id}}(\beta_{1,1}^0, \beta_{1,2}^0) &= t_g(\beta_{1,1}^0, \beta_{1,2}^0) \\
\begin{bmatrix}
\beta_{1,1}^0 \ \beta_{1,2}^0 \ 1
\end{bmatrix}
\Omega_{\text{Id}}
\begin{bmatrix}
\beta_{1,1}^0 \\
\beta_{1,2}^0 \\
1
\end{bmatrix}  
&= 
\begin{bmatrix}
\beta_{1,1}^0 \ \beta_{1,2}^0 \ 1
\end{bmatrix}
\Omega_g
\begin{bmatrix}
\beta_{1,1}^0 \\
\beta_{1,2}^0 \\
1
\end{bmatrix} \\
\begin{bmatrix}
\beta_{1,1}^0 \ \beta_{1,2}^0 \ 1
\end{bmatrix}
(\Omega_{\text{Id}} - \Omega_g)
\begin{bmatrix}
\beta_{1,1}^0 \\
\beta_{1,2}^0 \\
1
\end{bmatrix}
&= 0.
\end{aligned} 
$$
This is the matrix representation of a conic section (it can be degenerate such that the solution could be a point, two parallel lines, etc.). We can classify the general type of conic section (circle, ellipse, parabola, or hyperbola) by its discriminant $-4\Delta$, where $\Delta$ is the determinant of the upper-left $2\times2$ submatrix of $\Omega_{\text{Id}} - \Omega_{g}$. The $4$ is irrelevant in this context, so the discriminant is equivalent to $\mathbf{X}^T_{1,1} (A_\text{Id} - A_g) \left(\mathbf{X}_{1,2} \mathbf{X}^T_{1,1} - \mathbf{X}_{1,1} \mathbf{X}^T_{1,2}\right) (A_\text{Id} - A_g) \mathbf{X}_{1,2}$. The details for the classification can be found in Table \ref{conic}. This conic section partitions the $\mathbb{R}^2$ space so that depending on which side $[\beta_{1,1}^0, \beta_{1,2}^0]^T$ is relative to the conic section, $t_{\text{Id}}(\beta_{1,1}^0, \beta_{1,2}^0)$ will be greater or smaller than $t_{\text{g}}(\beta_{1,1}^0, \beta_{1,2}^0)$. 

\begin{table}[h!]
\centering
\caption{Conic section classification}
\label{conic}
\begin{tabular}{llll}
\hline
Conic Section & Ellipse & Parabola & Hyperbola\\
\midrule
Discriminant & $-\Delta < 0$ & $-\Delta = 0$ & $-\Delta > 0$\\
\hline
\end{tabular}
\caption*{\footnotesize$-\Delta_g = \mathbf{X}^T_{1,1} (A_\text{Id} - A_g) \left(\mathbf{X}_{1,2} \mathbf{X}^T_{1,1} - \mathbf{X}_{1,1} \mathbf{X}^T_{1,2}\right) (A_\text{Id} - A_g) \mathbf{X}_{1,2}$. As stated in the text, the $4$ is irrelevant in this context. The circle, which is a special kind of ellipse, corresponds to the case when $-\Delta_g < 0$ in addition to $\mathbf{X}^T_{1,1} (A_\text{Id} - A_g) \mathbf{X}_{1,2} = 0$ and $\mathbf{X}^T_{1,1} (A_\text{Id} - A_g) \mathbf{X}_{1,1} = \mathbf{X}^T_{1,2} (A_\text{Id} - A_g) \mathbf{X}_{1,2}$. Regardless of the value of the discriminant, the result could be a degenerate conic section.}
\end{table}

The $p$-value for any given value of $[\beta_{1,1}^0, \beta_{1,2}^0]^T$ is
$$
p(\beta_{1,1}^0, \beta_{1,2}^0) \equiv \frac{1}{M} \left(1 + \sum_{g \in \mathbf{G} \setminus \text{Id}} \mathbbm{1}\left[ t_{\text{Id}}(\beta_{1,1}^0, \beta_{1,2}^0) \leq t_g(\beta_{1,1}^0, \beta_{1,2}^0) \right] \right).
$$
The projected $p$-value function is equivalent to finding the max $p$-value over all other coefficients. For example, the projected $p$-value function for $\beta_{1,1}^0$ is
$$
p(\beta_{1,1}^0) = \max_{\beta_{1,2}^0} p(\beta_{1,1}^0, \beta_{1,2}^0).
$$

Unlike the scalar $\beta_1^0$ case, finding the exact confidence region for the vector $\beta_1^0$ case is difficult. It involves calculating the $p$-values in each region of the partitioned $\mathbb{R}^2$ space, then choosing the regions for which the $p$-value is greater than $\alpha$. As mentioned above, we present two methods.

The first method is fast grid search, but compared with traditional grid search methods, which would calculate the test and randomization statistics from start for each value of $[\beta_{1,1}^0, \beta_{1,2}^0]^T$, our method is an order faster because we only need to pay the fixed cost of calculating the coefficients of the second degree polynomial function once. After calculating these values, the marginal cost to calculate the test and randomization statistics for values of $[\beta_{1,1}^0, \beta_{1,2}^0]^T$ is very small. Calculating the $p$-value for any $[\beta_{1,1}^0, \beta_{1,2}^0]^T$ amounts to checking the proportion of $t_g(\beta_{1,1}^0, \beta_{1,2}^0)$ values greater than or equal to $t_{\text{Id}}(\beta_{1,1}^0, \beta_{1,2}^0)$. We fully outline the method in Algorithm \ref{gridSearchAlg}. To extract the level $1-\alpha$ valid confidence region(s), consider the values of $\beta_1^0$ for which the $p$-value is greater than $\alpha$, in addition to adjacent values of $\beta_1^0$ where rejection occurred.

\begin{algorithm}
\caption{Algorithm to compute approximate $p$-value function for Wald test using \cite{Pouliot2024}}
\label{gridSearchAlg}
\begin{algorithmic}[1]
    \State \textbf{Initialize:} Empty list $W$ to store $\Omega_g$ \hfill \textit{O(1)}
    \State \textbf{Initialize:} Empty matrix $P$ with dimensions $|\{\beta_1^0\}| \times |\{\beta_2^0\}|$ to store $p$-values \hfill \textit{O(1)}
    \State Calculate $\Omega_{\text{Id}}$ \hfill \textit{O(n)}
    \For{$g$ in $\mathbf{G} \setminus \text{Id}$} \hfill \textit{O(M)}
        \State Calculate $\Omega_g$ and store in $W$ \hfill \textit{O(n)}
    \EndFor
    \For{$x$ in $\{\beta_1^0\}$}
        \For{$y$ in $\{\beta_2^0\}$}
            \State $P[x,y] \gets \frac{1}{M} \left(1 + \sum_{g \in \mathbf{G} \setminus \text{Id}} \mathbbm{1}[t_{\text{Id}}(x, y) \leq t_g(x, y)] \right)$ \hfill \textit{O(M)}
        \EndFor
    \EndFor
    \State \textbf{Output:} $P$, the $p$-values for the grid of $\beta_1$ values.
\end{algorithmic}
\end{algorithm}

The second method involved optimization, but it requires that all the conic sections be ellipses. This seems to be a reasonable assumption in practice. While this method is easily extended to Exact Wald-tests of arbitrary dimension, we present this method for the two-dimensional case.

Note that when $-\Delta_g < 0$, then $f_g(\beta_{1,1}^0, \beta_{1,2}^0) = 
\begin{bmatrix}
\beta_{1,1}^0 \ \beta_{1,2}^0 \ 1
\end{bmatrix}
(\Omega_{\text{Id}} - \Omega_g)
\begin{bmatrix}
\beta_{1,1}^0 \ \beta_{1,2}^0 \ 1
\end{bmatrix}^T
\leq 0
$
is a convex elliptical disk. We plot three such disks for visualization purposes.

In this toy example, we see 8 distinct regions (since for each ellipse, we can either be inside or outside of it). To construct an exact projected $p$-value curve, one method is create a axis-aligned bounding box for each region so that we can easy project each region onto the $\beta{1,1}^0$ and $\beta_{1,2}^0$ axes. 

One could theoretically construct a bounding box for each region by solving a quadratically constrained problem. For instance, the optimization problem find the bounding box for region is
$$
\begin{aligned}
    \max_{(\beta_{1,1}^0, \beta_{1,2}^0)} &\beta_{1,1}^0 \\
    \text{s.t.} \quad &f_1(\beta_{1,1}^0, \beta_{1,2}^0) \leq 0 \\
    &f_2(\beta_{1,1}^0, \beta_{1,2}^0) \leq 0 \\
    &f_3(\beta_{1,1}^0, \beta_{1,2}^0) > 0 \\
\end{aligned}
$$
Clearly, however, the quadratic constraints are not convex, so finding the maximum value of $\beta_{1,1}^0$ subject to the constraints will not be easy. To solve this problem, one could remove the constraint that $f_3(\beta_{1,1}^0, \beta_{1,2}^0) > 0$. Then, instead of finding the bounding box for region $x$, we could easily find the bounding box for region $x \cup y$. While this is a larger bounding box, this has no effect on the projected $p$-value curve because we overwrite the projected $p$-value defined by $x \cup y$ with the projected $p$-value defined by $y$.

If we were interested in the projected confidence intervals for a particular level of significance, notice that there is no need to compute the entire projected $p$-value curve for $\beta_{1,1}^0$ and $\beta_{1,2}^0$. Only the portion of the $p$-value curve for which the $p$-value is greater than our level of significance, as that is the interval for which we fail to reject the null hypothesis. Therefore, we can modify the optimization problem to be a mixed integer quadratically constrained problem.

$$
\begin{aligned}
    \max_{(\beta_{1,1}^0, \beta_{1,2}^0)} &\beta_{1,1}^0 \\
    \text{s.t.} \quad &f_1(\beta_{1,1}^0, \beta_{1,2}^0) \leq M(1-z_1) \\
    &f_2(\beta_{1,1}^0, \beta_{1,2}^0) \leq M(1-z_2) \\
    &f_3(\beta_{1,1}^0, \beta_{1,2}^0) \leq M(1-z_3) \\
    &z_1 + z_2 + z_3 \geq 1\\
    &z_i \in \{0, 1\}
\end{aligned}
$$

Note that we added a constraint that the solution needs to be in a certain number of ellipses, that corresponds to our level of significance $\alpha$. These optimization programs can be solved using modern solvers.

\section{Applications to Other Tests}\label{otherTests}

The purpose of this paper to develop a general procedure to calculate the exact confidence intervals for permutation tests. To that end, we demonstrate that this method also extends to other tests in this section.

\subsection{Test Developed by \cite{d2024robust}}

\cite{d2024robust} also develop a heteroskedasticity-robust, Wald-like test. Their test, however, only handles discrete nuisance covariates, so consider the setup in Section \ref{onesidedPouliot} where $\mathbf{X}_2$ is composed of discrete covariates. We are primarily focused on the structure of the test and randomization statistic, so read \cite{d2024robust} for details about the inner workings of the test. The randomization statistic that they define is
$$
\mathcal{W}^{\pi} = g(W, (y - X\beta_0)_{\pi}),
$$
where
$$
g(W, v) = v' \Tilde{X}(\Tilde{X}' \Sigma(v) \Tilde{X})^{-1} \Tilde{X}' v;
$$
$W = [X, Z]$; $\Tilde{X}$ is a matrix derived from $X$; $v \in \mathbb{R}^{n \times 1}$; $\Sigma(v) \in \mathbb{R}^{n \times n}$ is a diagonal matrix with $(i,i)$ element equal to $(Dv)_i^2$; $D \in \mathbb{R}^{n \times n}$; and $\pi \in \mathbb{R}^{n \times n}$ is a permutation matrix. Like with the tests in \cite{Pouliot2024}, the test statistic corresponds to the case when $\pi = \text{Id}$. Notice that 
$$
\begin{aligned}
    \Tilde{X}' \Sigma(v) \Tilde{X} &= \frac{1}{n} \sum_{i=1}^{n} \Tilde{X}_i \Tilde{X}^T_i (Dv)^2_i.
\end{aligned}
$$
From this perspective, it is easy to see that the structure of the test statistic is similar to the over-identified instrumental-variable test statistic in Section \ref{waldPouliot} when $\beta_0$ is a scalar: $D$ and $\Tilde{X}$ are analogous to $\mathcal{Q}_3$ and $\mathcal{Q}_1 \mathbf{Z}$ respectively. We can follow the procedure outlined in Section \ref{waldPouliot} or \ref{jointValidPouliot} depending on whether $\beta^0$ is a scalar or vector to find the exact or approximate $p$-value curve and level $1-\alpha$ confidence region respectively.

\subsection{Test Developed by \cite{diciccio:2017aa}}

Another test that we touched upon in Section \ref{intro} is that of \cite{diciccio:2017aa}. \cite{diciccio:2017aa} develop a partial correlation test for specific components. Their test is not exact in finite samples, but is asymptotically valid. \cite{d2024robust} compare their method against the first randomization test in Section 3 of \cite{diciccio:2017aa}. That randomization statistic is 
$$
\begin{aligned}
    U_{\pi}(\mathbf{X}, \mathbf{Y}) &\equiv n \left(\hat{\beta}_{\pi} - \beta^0\right)^T \left( \hat{\Sigma}^{-1}_{\mathbf{XX}} \hat{\Omega}_{\pi}(\beta^0) \hat{\Sigma}^{-1}_{\mathbf{XX}} \right) \left(\hat{\beta}_{\pi} - \beta^0\right) \\
    &= \frac{n \left(\hat{\beta}_{\pi} - \beta^0\right)^T  \hat{\Omega}_{\pi}(\beta^0) \left(\hat{\beta}_{\pi} - \beta^0\right)}{\hat{\Sigma}^2_{\mathbf{XX}}},
\end{aligned}
$$
where $\hat{\beta}_{\pi}$ is the regression coefficients obtained from regressing $\pi Y$ on $X$; $\hat{\Sigma}^T_{\mathbf{XX}} = \frac{1}{n}\sum_i^n X^T_i X_i$ and $\hat{\Omega}_{\pi}(\beta^0) = \frac{1}{n} \sum_i^n ((\pi Y)_i - X^T_i \beta^0)^2 X_i X^T_i$.
Like with the other randomization tests, the test statistic corresponds to $\pi = \text{Id}$.

For simplicity, consider the case where $\beta^0 \equiv [\beta_1^0, \beta_2^0 ]^T \in \mathbb{R}^2$. Notice that $\hat{\Omega}(\beta^0)$ is similar in form to $\hat{\Sigma}_g(\beta_1^0)$ in the Wald statistic discussed in Section \ref{waldPouliot}. Then $\hat{\Omega}_{\pi}(\beta^0)$ can be rewritten as
$$
\begin{aligned}
    \hat{\Omega}_{\pi}(\beta^0) &= \begin{bmatrix}
    \hat{\omega}_{\pi}^{11}(\beta^0) & \hat{\omega}_{\pi}^{12}(\beta^0) \\
    \hat{\omega}_{\pi}^{21}(\beta^0) & \hat{\omega}_{\pi}^{22}(\beta^0) 
\end{bmatrix},
\end{aligned}
$$
where 
$$
\begin{aligned}
    \hat{\omega}_{\pi}^{vw} &= \frac{1}{n} \left(\pi \mathbf{Y} - \mathbf{X}^T \beta^0\right)^T \text{diag}[\mathbf{X}_v \circ \mathbf{X}_w] \left(\pi \mathbf{Y} - \mathbf{X}^T \beta^0\right) \\
    &= \frac{1}{n} \left(\pi \mathbf{Y} - \mathbf{X}_1^T \beta_1^0 - \mathbf{X}_2^T \beta_2^0\right)^T \text{diag}[\mathbf{X}_v \circ \mathbf{X}_w] \left(\pi \mathbf{Y} - \mathbf{X}_1^T \beta_1^0 - \mathbf{X}_2^T \beta_2^0 \beta^0\right) \\
    &= \frac{1}{n} \big(\mathbf{X}_1^T \Sigma_{vw} \mathbf{X}_1 (\beta_1^0)^2 + 2\mathbf{X}_1^T \Sigma_{vw} \mathbf{X}_2 (\beta_1^0 \ \beta_2^0) + \mathbf{X}_2^T \Sigma_{vw} \mathbf{X}_2 (\beta_2^0)^2  \\ 
    & \quad \quad -2\mathbf{Y}^T \pi^T \Sigma_{vw} \mathbf{X}_1 (\beta_1^0) -2\mathbf{Y}^T \pi^T \Sigma_{vw} \mathbf{X}_2 (\beta_2^0) + \mathbf{Y}^T \pi^T \Sigma_{vw} \pi \mathbf{Y} \big),
\end{aligned}
$$
and $\Sigma_{vw} = \text{diag}[\mathbf{X}_v \circ \mathbf{X}_w]$. Note that $\hat{\Omega}(\beta^0) \in \mathbb{R}^{2 \times 2}[\beta_1^0]$ and $(\hat{\beta} - \beta^0) \in \mathbb{R}^{2 \times 1}[\beta_1^0]$, so $U_{\pi}(\mathbf{X}, \mathbf{Y}) \in \mathbb{R}[\beta_1^0]$. Since $U_{\pi}(\mathbf{X}, \mathbf{Y})$ is just a polynomial, we can apply the same principles in Section \ref{jointValidPouliot} to construct jointly valid confidence regions. We will need to use grid search here, but we stress that our method is an order of magnitude faster than traditional grid search because we only calculate the coefficients of the polynomial for every permutation once rather than for each hypothesized value of $\beta^0$. Consequently, calculating the $p$-values and subsequent level $1-\alpha$ confidence interval for some region of $\beta^0$ will be much faster than traditional grid search.

%

\section{Conclusion}

The traditional method to produce confidence intervals in randomization-based inference in grid search, which is computationally expensive and leads to conservative confidence intervals as evidenced in \cite{d2024robust}. We show that many test and randomization statistics, including those where the variance estimator depends on the null hypothesis, are rational functions (if the variance estimator does not depend on the null hypothesis, the statistics reduce to a polynomial). We propose a method to use this structure to produce exact confidence intervals for a scalar $\beta_1^0$ for a broad class of test statistics. For cases where the null hypothesis $\beta_1^0$ is a vector, we use grid search; however, we show that by storing the coefficients of the function, our method is an order of magnitude faster than traditional grid search methods which require computing the test and randomization statistics from scratch for each null hypothesis value.

\begin{appendix}

\section{Two-Sided Hypothesis Test in \cite{Pouliot2024}}\label{pouliotTwoSided}

The randomization statistic for the two-sided hypothesis test for general regression analysis developed by \cite{Pouliot2024} is the absolute value of the one-sided randomization statistic defined in Section \ref{onesidedPouliot}:
$$
\begin{aligned}
    \left|T_g(\beta_1^0)\right| &\equiv \left|\frac{(\mathcal{Q}_1 \mathbf{X}_1)^\top g \mathbf{Y}}{\hat{\sigma}_g} - \frac{(\mathcal{Q}_1 \mathbf{X}_1)^\top g \mathbf{X}_1}{\hat{\sigma}_g} \beta_1^0\right| \\
    &= \left|m_g \beta_1^0 + b_g\right| \\
    &= \left|m_g\right| \left|\beta_1^0 + \frac{b_g}{m_g}\right|,
\end{aligned}
$$
where $m_g \neq 0$. Two absolute value functions that each have a single root will intersect at two, one, or an infinite number of points, but the latter two require $m_g$ to be equal, which has probability zero. If $m_{\text{Id}} \neq m_g$ \footnote{If $m_{\text{Id}} = m_g$ and $b_{\text{Id}} / m_{\text{Id}} = b_g / m_g$, then $\left|T_{\text{Id}}(\beta_1^0)\right| = \left|T_g(\beta_1^0)\right|$. If $m_{\text{Id}} = m_g$ and $b_{\text{Id}} / m_{\text{Id}} \neq b_g / m_g$, then there is one intersection point.}, then the two values of $\beta_1^0$ for which $\left|T_{\text{Id}}(\beta_1^0)\right|$ and $\left|T_g(\beta_1^0)\right|$ intersect can be found using the following formulas
$$
\dot{\beta}_1^0 = -\frac{b_{\text{Id}} + b_g}{m_{\text{Id}} + m_g}\text{,} \quad \text{ and } \quad \ddot{\beta}_1^0 = \frac{b_{\text{Id}} - b_g}{m_g - m_{\text{Id}}}.
$$
Using geometry, we know that if $|m_\text{Id}| > |m_g|$, then $|T_{\text{Id}}(\beta_1^0)| > |T_g(\beta_1^0)|$ when $\beta_1^0 \in (-\infty, \min\{\dot{\beta}_1^0, \ddot{\beta}_1^0\}) \cup (\max\{\dot{\beta}_1^0, \ddot{\beta}_1^0\}, \infty)$ and $|T_{\text{Id}}(\beta_1^0)| < |T_g(\beta_1^0)|$ when \\
$\beta_1^0 \in (\min\{\dot{\beta}_1^0, \ddot{\beta}_1^0\}, \max\{\dot{\beta}_1^0, \ddot{\beta}_1^0\})$. If $|m_\text{Id}| < |m_g|$, then the opposite case holds. The $p$-value function is
$$
p(\beta_1^0) = \frac{1}{M}\left(1 + \sum_{g \in \mathbf{G}_n \setminus \text{Id}} \mathbbm{1}\left[\left|T_{\text{Id}}(\beta_1^0)\right| \leq \left|T_g(\beta_1^0)\right|\right]\right).
$$
We can find the exact $p$-value curve using an algorithm similar to Algorithm \ref{PouliotTestAlg} and can easily extract the level $1-\alpha$ confidence interval(s) from the output. 

\end{appendix}

\bibliography{references}

\end{document}